\documentclass[11pt,twoside]{article}
\usepackage{asp2010}

\resetcounters
\begin{document}

\title{Andromeda and its Satellites - a Kinematic Perspective}
\author{Michelle L. M. Collins,$^1$ R. Mike Rich,$^2$ and Scott C. Chapman$^3$
\affil{$^1$MPIA, K\"onigstuhl 17, Heidelberg, 69117, Germany}
\affil{$^2$UCLA, Physics and Astronomy Building,
Los Angeles, CA 90095-1547, USA}
\affil{$^3$Institute of Astronomy, Madingley Rise, Cambridge, CB2 1ST, UK}}

\begin{abstract}
  Using spectroscopic data taken with Keck II DEIMOS by the Z-PAndAS team in
  M31, I present a comparison of the disc and satellite systems of M31 with
  those of the Galaxy. I discuss observed discrepancies between the masses and
  scale radii of M31 dwarf spheroidal galaxies of a given luminosity
  with those of the Milky Way. I also also present an analysis of the newly
  discovered M31 thick disc, which is measured to be hotter, more extended and
  thicker than that seen in the Milky Way.
\end{abstract}

\section{The Thick Disc in Andromeda}
Thick stellar discs are believed to represent an earlier epoch of galactic
evolution than thin stellar discs. Understanding the properties of these
structures is fundamental to our understanding of the evolution of large
spiral galaxies. Until recently, no such component was detected in Andromeda,
despite the belief that such structures are ubiquitous amongst spiral galaxies
\citep{dalcanton02}. We present a detection of a thick disc in the Andromeda
galaxy \citep{collins11a}, isolating the component by identifying a
significant population of stars that have velocities that lag behind the thin
stellar disc, and are inconsistent with the Andromedean halo.

We have been able to measure the scale lengths of the discs. We measure
h$_{r,thin}$= 7.3 kpc and h$_{r,thick}$= 8.0 kpc. Comparing these values with
thick disc systems in external galaxies (\citealt{yoachim06}, YD06) we
estimate the scale height of both components. We measure z$_{0,thin}$= 1.1 kpc
and z$_{0,thick}$= 2.8 kpc. We co-add their spectra to ascertain the
average metallicity of each component via the measurement of the equivalent
widths of the Ca II triplet (Fig. 1). The thick disc has an intermediate
metallicity of [Fe/H]= -1.0, more metal poor than the thin disc
([Fe/H]$=-0.7$) and more metal rich than the halo ([Fe/H]$=-1.3$).

\begin{figure}
\begin{center}
\includegraphics[angle=0,width=0.28\hsize]{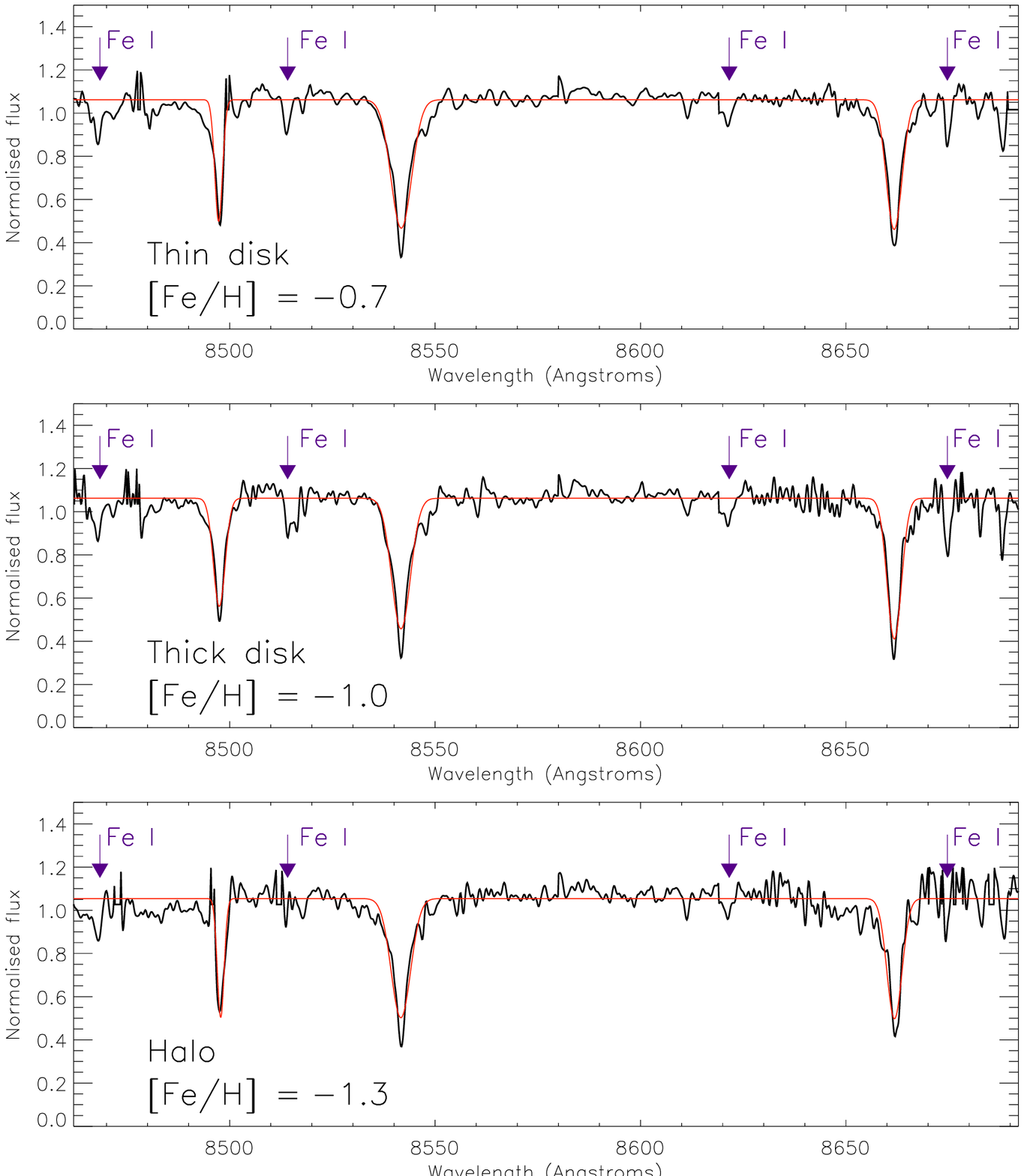}
\includegraphics[angle=0,width=0.3\hsize]{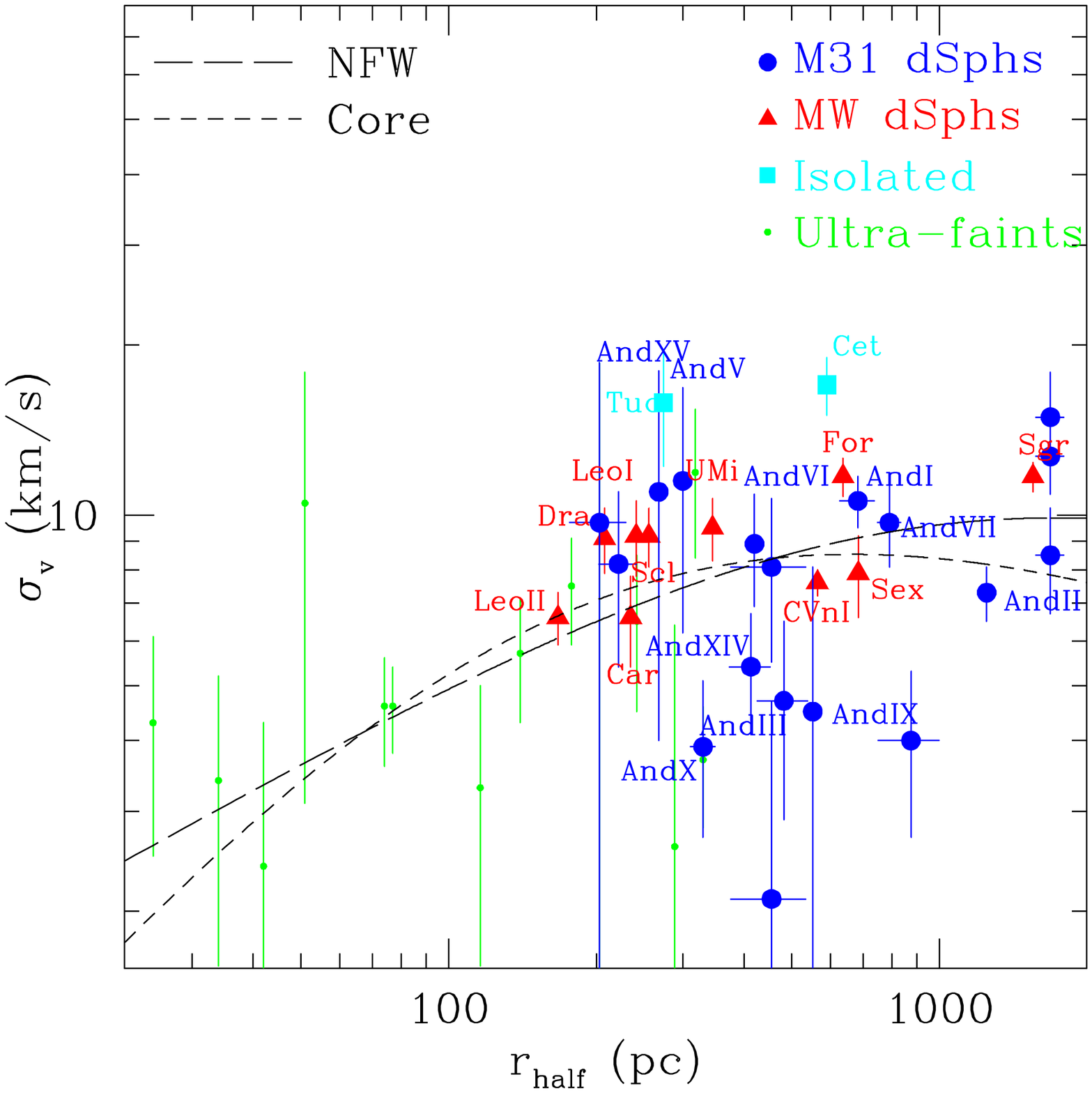}
\includegraphics[angle=0,width=0.3\hsize]{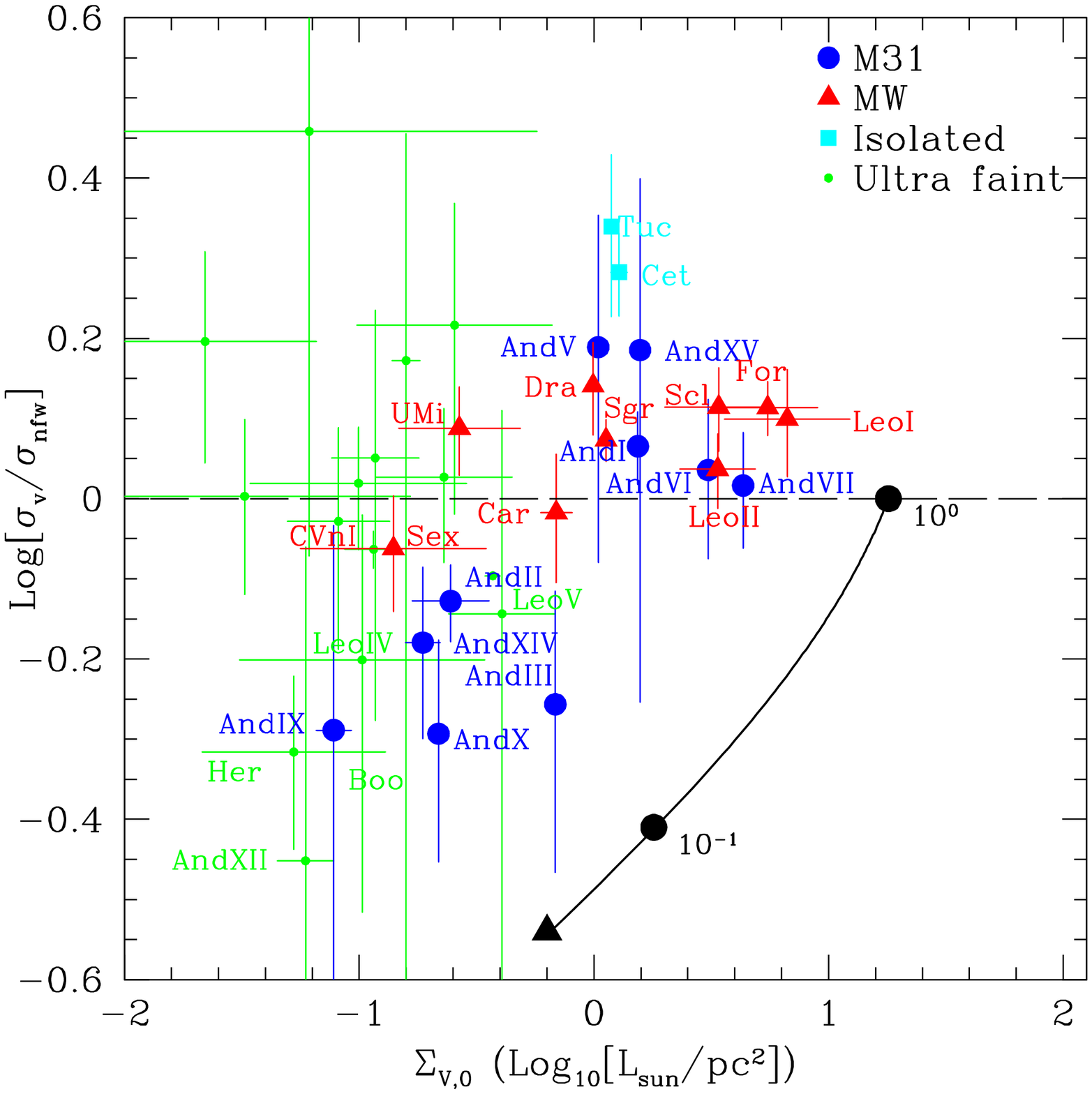}
\caption{{\bf Left: }Composite spectra for the thin (top), thick disc
  (middle) and halo (bottom) populations in M31. These are chemically
  distinct, with the thick disc having a metallicity of
  [Fe/H]$=-1.0$. {\bf Centre: }Velocity dispersion vs. half
light radius for MW (red) M31 (blue) and isolated (light blue) dSphs. The best
fit mass profiles of Walker et al. (2009) are overlaid. A number of M31 dSphs
have lower velocity dispersions than expected. {\bf
  Right: }Ratio of the measured : expected velocity dispersion from W09 NFW
profile for Local Group dSphs vs. surface brightness. The M31 outliers
are seen in the lower left corner. A tidal track from Penarrubia et al. (2008)
representing a dSph that has been stripped of 90 \% of its mass is
overplotted, and matches the observed trend.}
\end{center}
\end{figure}

\section{The Dwarf Spheroidals of Andromeda}

Dwarf spheroidal galaxies (dSphs) represent the smallest scales on which we
are able to detect dark matter. 27 dSphs have been identified in the M31 halo,
and while the relationship between the size and luminosity of these objects is
consistent with the MW dSphs \citep{brasseur11}, there are a number
that are more extended than their MW counterparts (\citealt{mcconnachie06b,
  richardson11}, Fig. 1).

Work by Walker et al. (2009) demonstrated that the MW dSphs were consistent
with having formed in halos with identical dark matter profiles, suggesting
that the evolution of the dark matter within these objects is the same,
regardless of the size of the luminous component. Our group has been measuring
the kinematics of the M31 dSphs
\citep{chapman05,chapman07,collins10,collins11b}. We have discovered that
these objects are not consistent with the MW population, as a number of them
have lower velocity dispersions for a given half light radius when compared to
the MW. These outliers are also typically the more extended M31 dSphs
(Fig. 1), consistent with the effects of increased tidal stress from a more
massive M31 disc \citep{penarrubia08b,penarrubia10}. Recent work by the SPLASH
collaboration \citep{tollerud12} also note some outliers in their survey, but
conclude that overall, the difference between the masses of MW and M31 dSphs
are not statistically significant.

\bibliography{michelle}

\end{document}